\pdfoutput=1
\documentclass[journal=jacsat,manuscript=article]{achemso}
\setkeys{acs}{usetitle=true}

\usepackage{graphicx}
\usepackage[space]{grffile}
\usepackage{latexsym}
\usepackage{textcomp}
\usepackage{longtable}
\usepackage{multirow,booktabs}
\usepackage{amsfonts,amsmath,amssymb}
\usepackage{natbib}
\usepackage{color}
\usepackage{array}
\usepackage{url}
\usepackage[utf8]{inputenc}
\usepackage{hyperref}
\hypersetup{colorlinks=false,pdfborder={0 0 0}}
\newif\iflatexml\latexmlfalse
\usepackage[version=3]{mhchem} 
\usepackage[T1]{fontenc}

\usepackage{tikz}
\usepackage{gb4e}
\noautomath

\usepackage[final]{changes}

\newcolumntype{L}{>{\centering\arraybackslash}m{2cm}}
\newcolumntype{K}{>{\centering\arraybackslash}m{3cm}}

\author{Jennifer N. Wei}
\affiliation{Department of Chemistry and Chemical Biology, Harvard University, Cambridge MA 02138, USA}
\author{David Duvenaud}
\affiliation{Department of Computer Science, Harvard University, Cambridge MA 02138, USA}
\author{Al{\'a}n Aspuru-Guzik}
\affiliation{Department of Chemistry and Chemical Biology, Harvard University, Cambridge MA 02138, USA}
\email{aspuru@chemistry.harvard.edu}

\title[]{Neural networks for the prediction of organic chemistry reactions}

\begin{document}

\begin{abstract}
Reaction prediction remains one of the major challenges for organic chemistry, and is a prerequisite for efficient synthetic planning. It is desirable to develop algorithms that, like humans, "learn" from being exposed to examples of the application of the rules of organic chemistry. We explore the use of neural networks for predicting reaction types, using a new reaction fingerprinting method.  We combine this predictor with SMARTS transformations to build a system which, given a set of reagents and reactants, predicts the likely products.  We test this method on problems from a popular organic chemistry textbook.
\end{abstract}

\section{Introduction}

To develop the intuition and understanding for predicting reactions, a human must take many semesters of organic chemistry and gather insight over several years of lab experience. Over the past 40 years, various algorithms have been developed to assist with synthetic design, reaction prediction, and starting material selection\cite{Todd_2005,Szymkuc_2016}. LHASA was the first of these algorithms to aid in developing retrosynthetic pathways\cite{Corey1971Centenary}. This algorithm required over a decade of effort to encode the necessary subroutines to account for the various subtleties of retrosynthesis such as functional group identification, polycyclic group handling, relative protecting group reactivity, and functional group based transforms \cite{corey1972techniques,corey1975general,corey1985computer,corey1972computer}. 

In the late 1980s to the early 1990s, new algorithms for synthetic design and reaction prediction were developed. CAMEO\cite{jorgensen1990cameo}, a reaction predicting code, used subroutines specialized for each reaction type, expanding to include reaction conditions in its analysis. EROS\cite{gasteiger1987new} identified leading structures for retrosynthesis by using bond polarity, electronegativity across the molecule, and the resonance effect to identify the most reactive bond. SOPHIA\cite{Satoh_1996} was developed to predict reaction outcomes with minimal user input; this algorithm would guess the correct reaction type subroutine to use by identifying important groups in the reactants; once the reactant type was identified, product ratios would be estimated for the resulting products. SOPHIA was followed by the KOSP algorithm, and uses the same database to predict retrosynthetic targets\cite{Satoh_1999}. Other methods generated rules based on published reactions, and uses these transformations when designing a retrosynthetic pathway\cite{ChemPlanner,Gelernter_1990}. Some methods encoded expert rules in the form of electron flow diagrams \cite{Baldi2009NoElectron,Chen2008SynthExplore}. Another group attempted to grasp the diversity of reactions by creating an algorithm that automatically searches for reaction mechanisms using atom mapping and substructure matching\cite{Law_2009}. 

While these algorithms have their subtle differences, all require a set of expert rules to predict reaction outcomes. Taking a more general approach, one group has encoded all of the reactions of the Beilstein database, creating a 'Network of Organic Chemistry'\cite{Gothard_2012,Szymkuc_2016}. By searching this network, synthetic pathways can be developed for any molecule similar enough to a molecule already in its database of 7 million reactions, identifying both one-pot reactions that do not require time-consuming purification of intermediate products\cite{Grzybowski_2009}, or full multistep reactions that account for the cost of the materials, labor, and safety of the reaction\cite{Szymkuc_2016}. Algorithms that use encoded expert rules or databases of published reactions are able to accurately predict chemistry for queries that match reactions in its knowledge base. However, such algorithms do not have the ability of a human organic chemist to predict the outcomes of previously unseen reactions. In order to predict the results of new reactions, the algorithm must have a way of connecting information from reactions that it has been trained upon to reactions that it has yet to encounter.

Another strategy of reaction prediction algorithm draws from principles of physical chemistry and first predicts the energy barrier of a reaction in order to predict its likelihood\cite{Zimmerman_2013,Wang_2014,Wang_2016,xu2009dynamics,Rappoport_2014,Socorro_2005}. Specific examples of reactions include the development of a nanoreactor for early Earth reactions\cite{Wang_2014,Wang_2016}, Heuristic Aided Quantum Chemistry \cite{Rappoport_2014}, and ROBIA\cite{socorro2005robia}, an algorithm for reaction prediction. While methods that are guided by quantum calculations have the potential to explore a wider range of reactions than the heuristic-based methods, these algorithms would require new calculations for each additional reaction family, and will be prohibitively costly over a large set of new reactions.

A third strategy for reaction prediction algorithms uses statistical machine learning. These methods can sometimes generalize or extrapolate to new examples, as in the recent examples of picture and handwriting identification \cite{He_2015,krizhevsky2012imagenet}, playing video games \cite{mnih2015human}, and most recently, playing Go \cite{silver2016mastering}. This last example is particularly interesting as Go is a complex board game with a search space of $10^{170}$, which is on the order of chemical space for medium sized molecules\cite{reymond2010chemical}.  SYNCHEM was one early effort in the application of machine learning methods to chemical predictions, which relied mostly on clustering similar reactions, and learning when reactions could be applied based on the presence of key functional groups\cite{Gelernter_1990}. 

Today, most machine learning approaches in reaction prediction use molecular descriptors to characterize the reactants in order to guess the outcome of the reaction. Such descriptors range from physical descriptors such as molecular weight, number of rings, or partial charge calculations to molecular fingerprints, a vector of bits or floats that represent the properties of the molecule. ReactionPredictor\cite{Kayala_2012, Kayala2011Learning} is an algorithm that first identifies potential electron sources and electron sinks in the reactant molecules based on atom and bond descriptors. Once identified, these sources and sinks are paired to generate possible reaction mechanisms. Finally, neural networks are used to determine the most likely combinations in order to predict the true mechanism. While this approach allows for the prediction of many reactions at the mechanistic level, many of the elementary organic chemistry reactions that are the building blocks of organic synthesis have complicated mechanisms, requiring several steps that would be costly for this algorithm to predict.

Many algorithms that predict properties of organic molecules use various types of fingerprints as the descriptor. Morgan fingerprints and extended circular fingerprints \cite{Morgan_1965,Rogers2010Extended} have been used to predict molecular properties such as HOMO-LUMO gaps\cite{Pyzer_Knapp_2015a}, protein-ligand binding affinity\cite{Ballester_2010}, drug toxicity levels\cite{Zhang_2006}, and even to predict synthetic accessibility\cite{Podolyan_2010}. Recently Duvenavud et al. applied graph neural networks\cite{duvenaud_convolutional_2015} to generate continuous molecular fingerprints directly from molecular graphs. This approach generalizes fingerprinting methods such as the ECFP by parameterizing the fingerprint generation method. These parameters can then be optimized for each prediction task, producing fingerprint features that are relevant for the task. Other fingerprinting methods that have been developed use the Coulomb matrix\cite{montavon2012learning}, radial distribution functions\cite{von2015fourier}, and atom pair descriptors\cite{Carhart_1985}. For classifying reactions, one group developed a fingerprint to represent a reaction by taking the difference between the sum of the fingerprints of the products and sum of the fingerprints of the reactants\cite{Schneider_2015}. A variety of fingerprinting methods were tested for the constituent fingerprints of the molecules.

\begin{figure}
\begin{center}
\includegraphics[width=0.9\columnwidth]{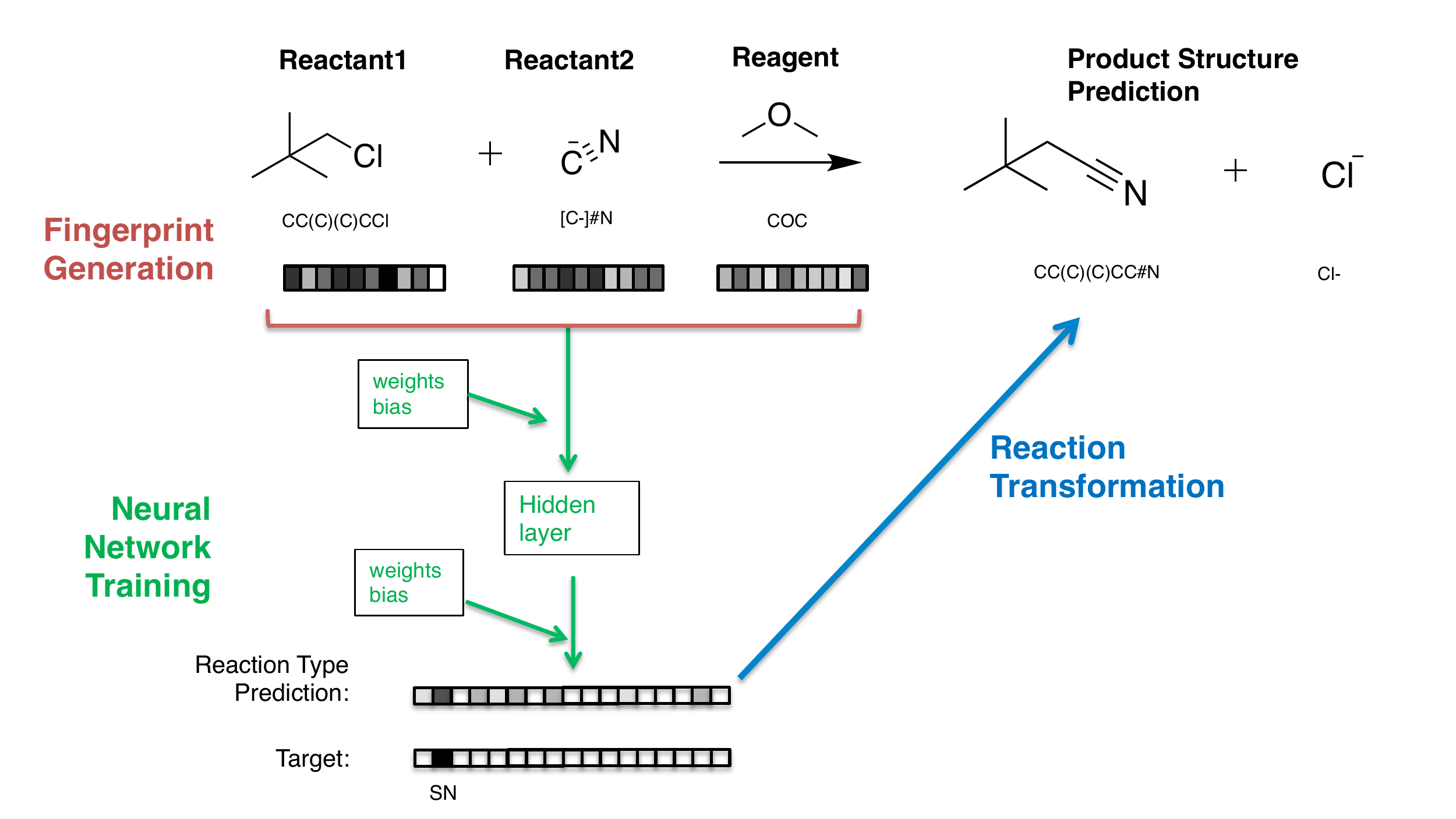}
\caption{\label{Fig:PredictionOverview} An overview of our method for predicting reaction type and products. A reaction fingerprint, made from concatenating the fingerprints of reactant and reagent molecules, is the inputs for a neural network that predicts the probability of 17 different reaction types, represented as a reaction type probability vector. The algorithm then predicts a product by applying a transformation that corresponds with the most probable reaction type to the reactants. In this work, we use a SMARTS transformation for the final step.}
\end{center}
\end{figure}

In this work, we apply fingerprinting methods, including neural molecular fingerprints, to predict organic chemistry reactions. Our algorithm predicts the most likely reaction type for a given set of reactants and reagents, using what it has learned from training examples. \deleted{To do so, we use a new kind of a reaction fingerprint, made from concatenating the fingerprints of the reactants and the reagents.} \added{These input molecules are described by concatenating the fingerprints of the reactants and the reagents; this concatenated fingerprint is then used as the input for a neural network to classify the reaction type. With information about the reaction type, we can make predictions about the product molecules. One simple approach for predicting product molecules from the reactant molecules, which we use in this work, is to apply a SMARTS transformation that describes the predicted reaction. Previously, sets of SMARTS transformations have been applied to produce large libraries of synthetically accessible compounds in the areas of molecular discovery\cite{Virshup_2013}, metabolic networks\cite{mann2013graph}, drug discovery\cite{Schuerer_2005}, and discovering one-pot reactions\cite{gothard2012rewiring}. In our algorithm, we use SMARTS transformation for targeted prediction of product molecules from reactants. However, this method can be replaced by any method that generates product molecule graphs from reactant molecule graphs.}  An overview of our method can be found in \ref{Fig:PredictionOverview}, and is explained in further detail in the Prediction Methods section.

We show the results of our prediction method on 16 basic reactions of alkylhalides and alkenes, some of the first reactions taught to organic chemistry students in many textbooks\cite{wade2013organic}.  The training and validation reactions were generated by applying simple SMARTS transformations to alkenes and alkylhalides. While we limit our initial exploration to aliphatic, non-stereospecific molecules, our method can easily be applied a wider span of organic chemical space with enough example reactions. The algorithm can also be expanded to include experimental conditions such as reaction temperature and time. With additional adjustments and a larger library of training data, our algorithm will be able to predict multistep reactions, and eventually, become a module in a larger machine-learning system for suggesting retrosynthetic pathways for complex molecules.

\section{Results and Discussion}

\subsection{Performance on cross-validation set}

We created a dataset of reactions of four alkylhalide reactions and twelve alkene reactions; further details on the construction of the dataset can be found in the Methods section. Our training set comprised of 3400 reactions from this dataset, and the test set comprised of 17,000 reactions; both the training set and the test set were balanced across reaction types. During optimization on the training set, k-fold cross-validation was used to help tune the parameters of the neural net. Table 1 reports the cross-entropy score and the accuracy of the baseline and fingerprinting methods on this test set. Here the accuracy is defined by the percentage of matching indices of maximum values in the predicted probability vector and the target probability vector for each reaction.

\begin{table}
    \centering
    \label{tab:cross_val_results}
    \begin{tabular}{ |K|K|L|L|L|L|}
         \hline
         \textbf{Fingerprint Method} & \textbf{Fingerprint length} & \textbf{Train NLL} & \textbf{Train Accuracy} & \textbf{Test NLL} &  \textbf{Test Accuracy} \\ 
        \hline
        Baseline & 51 & 0.2727 & 78.8\% & 2.5573 & 24.7\%  \\ 
        \hline
        Morgan  & 891 & 0.0971 & 86.0\%  & 0.1792 & 84.5\%  \\ 
        Neural  &  181 & 0.0976 & 86.0\% & 0.1340 & 85.7\% \\ 
        \hline
    \end{tabular} 
    \caption{Accuracy and Negative Log Likelihood (NLL) Error of fingerprint and baseline methods} 
\end{table}  

\begin{figure}[ht!]
\begin{center}
\includegraphics[trim=80 180 80 70,clip,width=0.9\textwidth]{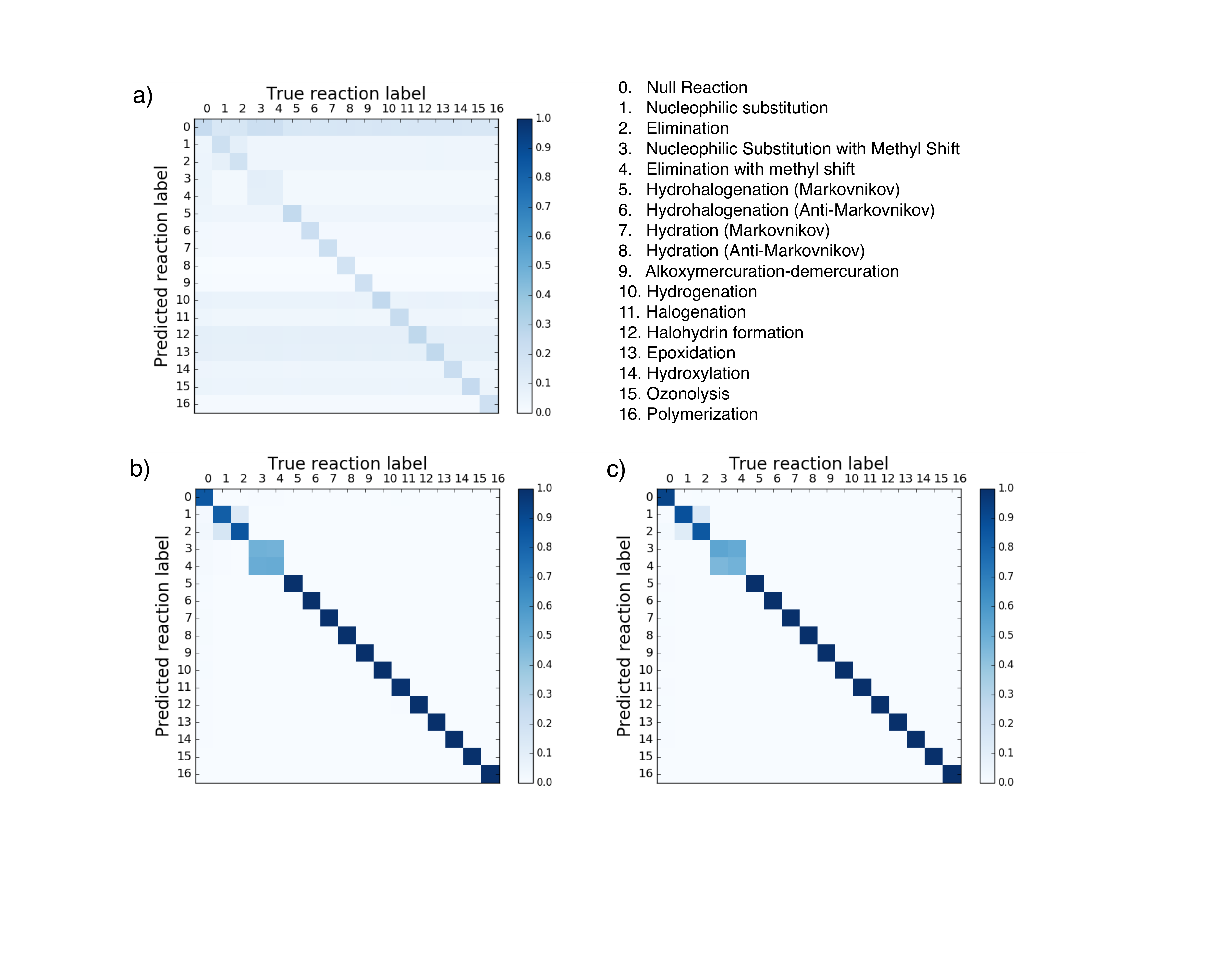}
\caption{\label{fig:conf_mat} Cross validation results for a) Baseline fingerprint, b) Morgan reaction fingerprint, and c) neural reaction fingerprint. \added{A confusion matrix shows the average predicted probability for each reaction type.} In these confusion matrices, the predicted reaction type is represented on the vertical axis, and the correct reaction type is represented on the horizantal axis. These figures were generated based on code from Schneider et al.\cite{Schneider_2015}.
}
\end{center}
\end{figure}

Figure \ref{fig:conf_mat} shows the confusion matrices for the baseline, neural, and Morgan fingerprinting methods respectively. The confusion matrices for the Morgan and neural fingerprints show that the predicted reaction type and the true reaction type correspond almost perfectly, with few mismatches. The only exceptions are in the predictions for reaction types 3 and 4, corresponding to nucleophilic substitution reaction with a methyl shift and the elimination reaction with a methyl shift. As described in the methods section, these reactions are assumed to occur together, so they are each assigned probabilities of 50\% in the training set. As a result, the algorithm cannot distinguish these reaction type and the result on the confusion matrix is a 2x2 square. For the baseline method, the first reaction type, the 'NR' classification, is often over predicted, with some additional overgeneralization of some other reaction type as shown by the horizontal bands.

\subsection{Performance on predicting reaction type of exam questions}

\added{Kayala et al.\cite{Kayala_2012} had previously employed organic textbook questions both as the training set and as the validation set for their algorithm, reporting 95.7\% accuracy on their training set. We similarly decided to test our algorithm on a set of textbook questions.} \deleted{To challenge our algorithm, we tested the performance on textbook problems that an organic chemistry student would see.} We selected problems 8-47 and 8-48 from the Wade 6th edition organic chemistry textbook shown below in Figure \ref{fig:Wade_problems_copy}.\cite{wade2013organic} The reagents listed in each problem were assigned as secondary reactants or reagents so that they matched the training set. For all prediction methods, our networks were first trained on the training set of generated reactions, using the same hyperparameters found by the cross-validation search. \added{The similarity of the exam questions to the training set was determined by measuring the Tanimoto\cite{Bajusz_2015} distance of the fingerprints of the reactant and reagent molecules in each reactant set. The average Tanimoto score between the training set reactants and reagents and the exam set reactants and reagents is 0.433, and the highest Tanimoto score oberved between exam questions and training questions was 1.00 on 8-48c and 0.941 on 8-47a. This indicates that 8-48c was one of the training set examples. Table SI.1  show more detailed results for this Tanimoto analysis.} \deleted{Except where the test questions and the training set overlapped, the algorithm had not seen any of these questions.}

\begin{figure}
\begin{center}
\includegraphics[width=0.9\textwidth]{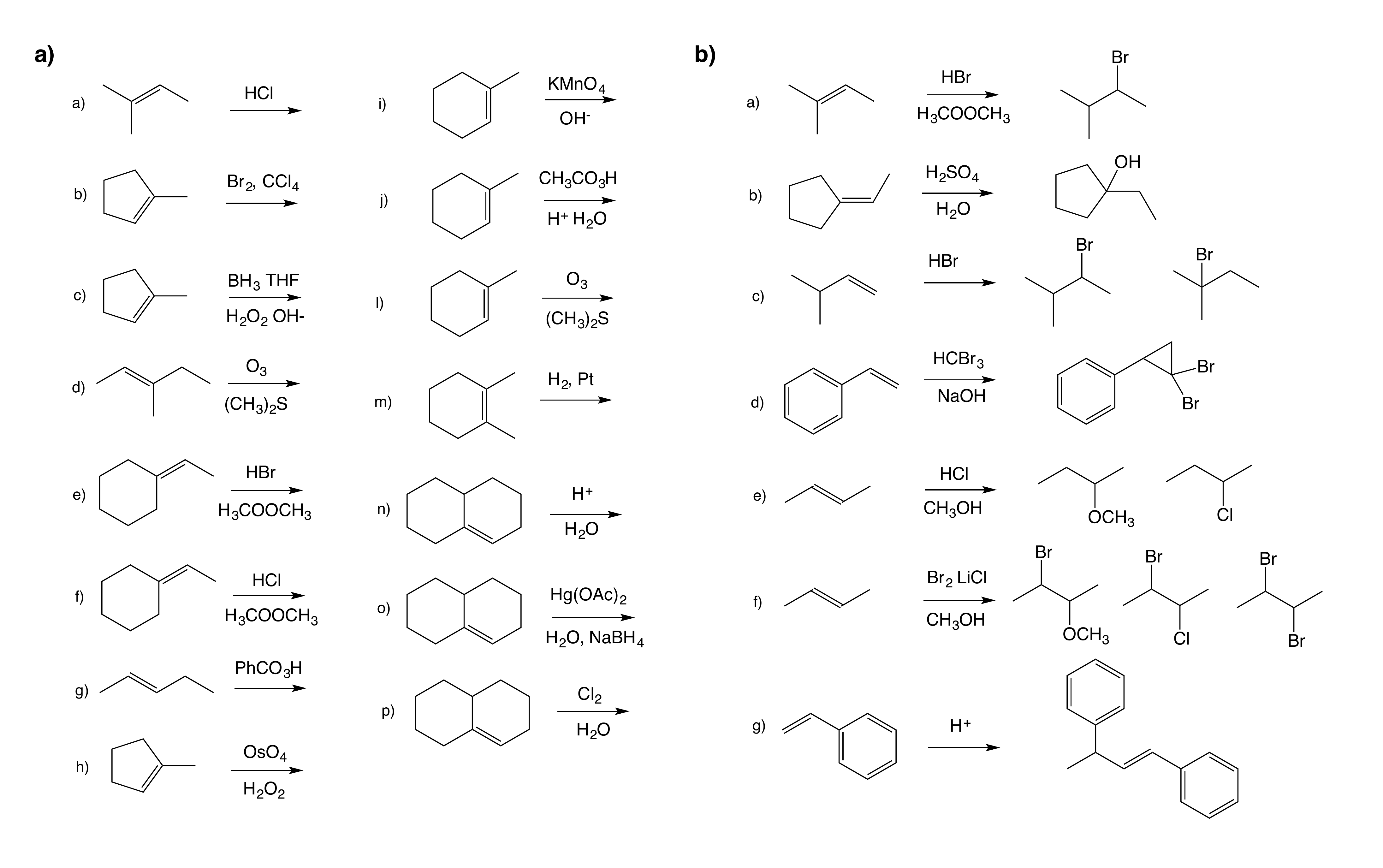}
\caption{\label{fig:Wade_problems_copy} Wade problems a) 8-47 and b) 8-48}
\end{center}
\end{figure}

\begin{figure}
\begin{center}
\includegraphics[trim=20 450 165 140,clip,width=0.9\textwidth]{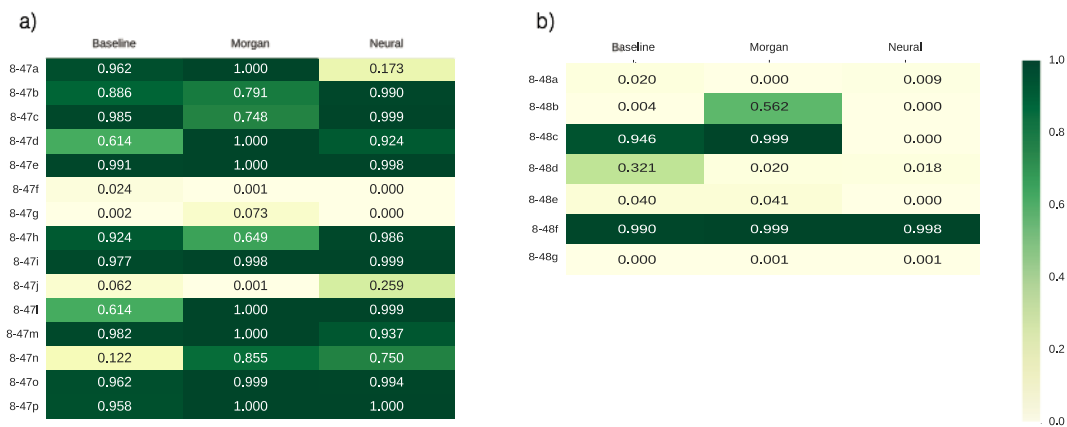}
\caption{\label{fig:Wade_results} Prediction results for a) Wade Problem 8-47 and b) Wade Problem 8-48, as displayed by estimated probability of correct reaction type. Darker (greener) colors represent a higher predicted probability. Note the large amount of correct predictions in 8-47}
\end{center}
\end{figure}

For each problem, the algorithm determined the reaction type in our set that best matched the answer. If the reaction in the answer key did not match any of our reaction types, the algorithm designated the reaction as a null reaction. The higher the probability the algorithm assigned for each reaction type, the more certainty the algorithm has in its prediction. These probabilities are reported below in Figure \ref{fig:Wade_results}, color-coded with green for probability and yellow/white for low probability. 

In problem 8-47, the Morgan fingerprint algorithm had the best performance with 12 of the 15 correct answers, followed by the neural fingerprint algorithm and the baseline method, both of which had 11 out of 15 correct answers. Both the Morgan fingerprint algorithm and the neural fingerprint algorithm predicted the correct answers with higher probability than the baseline method. Several of the problems contained rings, which weren't included in the original training set. Many of these reactions were predicted correctly by the Morgan and neural fingerprint algorithm, but not by the baseline algorithm. This suggests that both Morgan and neural fingerprint algorithms were able to extrapolate the correct reactivity patterns to reactants with rings.

In problem 8-48, students are asked to suggest mechanisms for reactions given the both the reactants and the products. To match the input format of our algorithm, we did not provide the algorithm any information about the products even though it disadvantaged our algorithm. All methods had much greater difficulty with this set of problems possibly because these problems introduced aromatic rings, which the algorithm may have had difficulty distinguishing from double bonds.

\subsection{Performance on Product Prediction}

\begin{figure}
\begin{center}
\includegraphics[ trim=120 180 120 100, clip,width=0.9\textwidth]{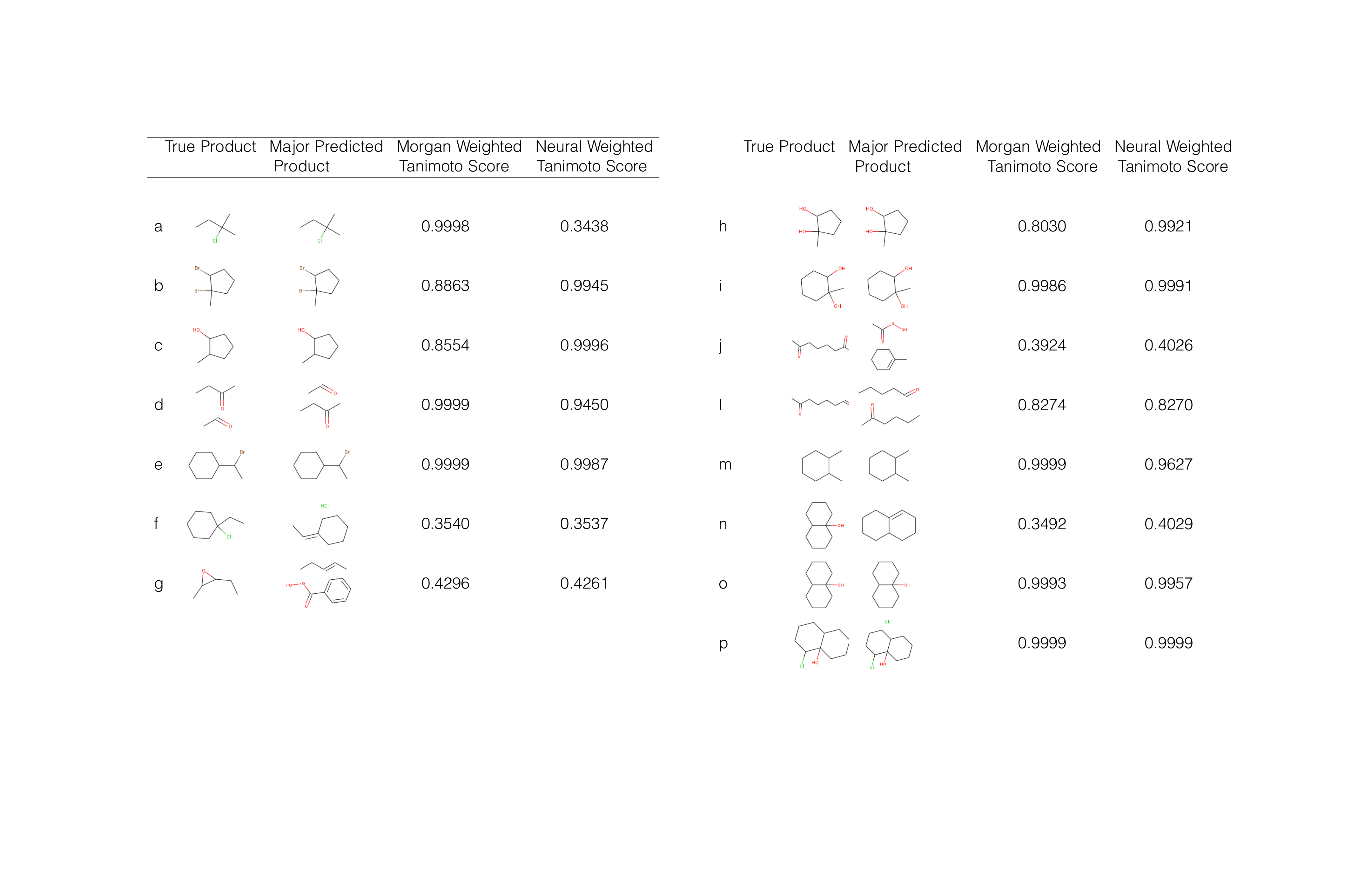}
\caption{\label{Fig:prod_pred_Wade847} Product predictions for Wade 8-47 questions, with Tanimoto score. The true product is the product as defined by the answer key. The major predicted product shows the product of the reaction type with the highest probability according to the Morgan fingerprint algorithm's result. The Morgan weighted score and the neural weighted score are calculated by taking an average of the Tanimoto scores over all the predicted products weighted by the probability of that reaction type which generated that product.}
\end{center}
\end{figure}

Once a reaction type has been assigned for a given problem by our algorithm, we can use the information to help us predict our products. In this study, we chose to naively use this information by applying a SMARTS transformation that matched the predicted reaction type to generate products from reactants. Figure \ref{Fig:prod_pred_Wade847} shows the results of this product prediction method using Morgan reaction fingerprints and neural reaction fingerprints on problem 8-47 of the Wade textbook, analyzed in the previous section. For all suggested reaction types, the SMARTS transformation was applied to the reactants given by the problem. If the SMARTS transformation for that reaction type was unable to proceed due to a mismatch between the given reactants and the template of the SMARTS transformation, then the reactants were returned as the predicted product instead. 

A product prediction score was also assigned for each prediction method. For each reaction, the Tanimoto score\cite{Bajusz_2015} was calculated between the Morgan fingerprint of the true product and the Morgan fingerprint of the predicted product for each reaction type, following the same applicability rules described above. The overall product prediction score is defined as average of these Tanimoto scores for each reaction type, weighted by the probability of each reaction type as given by the probability vector. The scores for each question are given in Fig. \ref{Fig:prod_pred_Wade847}.

The Morgan fingerprint algorithm is able to predict 8 of the 15 products correctly, the neural fingerprint algorithm is able to predict 7 of the 15 products correctly. The average Tanimoto score for the products predicted by the Morgan fingerprint algorithm compared to the true products was 0.793 and the average Tanimoto score between the true products and the neural fingerprint algorithm products was 0.776. In general, if the algorithm predicted the reaction type correctly with high certainty, the product was also predicted correctly and the weighted Tanimoto score was high, however, this was not the case for all problems correctly predicted by the algorithm.

The main limitation in the algorithm's ability to predict products despite predicting the reaction type correctly is the capability of the SMARTS transformation to accurately describe the transformation of the reaction type for all input reactants. While some special measures were taken in the code of these reactions to handle some common regiochemistry considerations, such as Markovnikov orientation, it was not enough to account for all of the variations of transformations seen in the sampled textbook questions. Future versions of this algorithm will require an algorithm better than encoded SMARTS transformations to generate the products from the reactant molecules. 

\section{Conclusion}

Using our fingerprint-based neural network algorithm, we were able to identify the correct reaction type for most reactions in our scope of alkene and alkylhalide reactions, given only the reactants and reagents as inputs. We achieved an accuracy of 85\% of our test reactions and 80\% of selected textbook questions. With this prediction of the reaction type, the algorithm was further able to guess the structure of the product for a little more than half of the problems. The main limitation in the prediction of the product structure was due to the limitations of the SMARTS transformation to describe the mechanism of the reaction type completely.

\added{While previously developed machine learning algorithms are also able to predict the products of these reactions with similar or better accuracy\cite{Kayala_2012}, the structure of our algorithm allows for greater flexibillity. Our algorithm is able to learn the probabilities of a range of reaction types. To expand the scope of our algorithm to new reaction types, we would not need to encode new rules, nor would we need to account for the varying number of steps in the mechanism of the reaction; we would just need to add the additional reactions to the training set. The simplicity of our reaction fingerprinting algorithm allows for rapid expansion of our predictive capabilities given a larger dataset of well-curated reactions\cite{Szymkuc_2016,ChemPlanner}. Using datasets of experimentally published reactions, we can also expand our algorithm to account for the reaction conditions in its predictions, and later, predict the correct reaction conditions.}

\added{This paper represents a step towards the goal of developing a machine learning algorithm for automatic synthesis planning for organic molecules. Once we have an algorithm that can predict the reactions that are possible from its starting materials, we can begin to use the algorithm to string these reactions together to develop a multistep synthetic pathway. This pathway prediction can be further optimized to account for reaction conditions, cost of materials, fewest number of reaction steps and other factors to find the ideal synthetic pathway. Using neural networks helps the algorithm to identify important features from the reactant molecules structure in order to classify new reaction types.}

\deleted{The simplicity of our reaction fingerprinting algorithm allows for rapid expansion of our predictive capabilities given a larger dataset of well-curated reactions\cite{Szymkuc_2016,ChemPlanner}. The inclusion of experimental conditions into the dataset would allow our algorithm to predict full reactions. \
Once individual elementary reaction steps can be identified by the reactants, and the products predicted, we can predict multi-step reactions, and eventually, incorporate this program into a larger machine learning system whose goal would be that of predicting retrosynthetic pathways for synthesizing any reasonable molecule returned by a high-throughput virtual search.}

\section{Methods}

\subsection{ Dataset Generation  }

The data set of reactions was developed as follows: A library of all alkanes containing 10 carbon atoms or fewer was constructed. To each alkane, a single functional group was added, either a double bond or a halide (Br, I, Cl). Duplicates were removed from this set to make the substrate library. Sixteen different reactions were considered, 4 reactions for alkylhalides and 12 reactions for alkenes. Reactions resulting in methyl shifts, or resulting in Markovnikov or anti-Markovnikov product were considered as separate reaction types. Each reaction is associated with a list of secondary reactants and reagents, as well as a SMARTS transformation to generate the product structures from the reactants. 

To generate the reactions, every substrate in the library was combined with every possible set of secondary reactants and reagents. Those combinations that matched the reaction conditions set by our expert rules, were assigned a reaction type. \added{If none of the reaction conditions were met, the reaction was designated a 'Null Reaction' or NR for short. We generated a target probability vector to reflect this reaction type assignment with a one-hot encoding; that is, the index in the probability vector that matches the assigned reaction type had a probability of 1, and all other reaction types had a probability of 0.} \deleted{A target probability vector was generated to reflect the assigned reaction type(s), This vector is a one-hot encoded vector, with 100\% on the index representing the reaction type.} The notable exception to this rule was for the elimination and substitution reactions involving methyl shifts for bulky alkylhalides; these reactions were assumed to occur together, and so 50\% was assigned to each index corresponding to these reactions. Products were generated using the SMARTS transformation associated with the reaction type with the two reactants as inputs. Substrates that did not match the reaction conditions were designated 'null reactions' (NR), indicating that the final result of the reaction is unknown. RDKit \cite{landrum2006rdkit} was used to handle the requirements and the SMARTS transformation. A total of 1,277,329 alkyhalide and alkene reactions were generated. A target reaction probability vector was generated for each reaction. 

\subsection{Prediction Methods }

\added{As outlined in Figure \ref{Fig:PredictionOverview}, to predict the reaction outcomes of a given query, we first predict the probability of each reaction type in our dataset occurring, then we apply SMARTS transformations associated with each reaction.} The reaction probability vector\added{, i.e. the vector encoding the probability of all reactions,} was predicted using a neural network with reaction fingerprints as the inputs. This reaction fingerprint was formed as a concatenation of the molecular fingerprints of the substrate (Reactant1), the secondary reactant (Reactant2) and the reagent. Both the Morgan fingerprint method\added{, in particular the extended-connectivity circular fingerprint (ECFP),} and the neural fingerprint method were tested for generating the molecular fingerprints. \added{A Morgan circular fingerprint hashes the features of a molecule for each atom at each layer into a bit vector. Each layer considers atoms in the neighborhood of the starting atom that are less than the maximum distance assigned for that layer. Information from previous layers is incorporated into later layers, until highest layer, e.g. maximum bond length radius, is reached\cite{Rogers2010Extended}. A neural fingerprint also records atomic features at all neighborhood layers, but instead of using a hash function to record features, uses a convolutional neural network, thus creating a fingerprint with differentiable weights. Further discussion about circular fingerprints and neural fingerprints can be found in Duvenaud et al\cite{duvenaud_convolutional_2015}.} The circular fingerprints were generated with RDKit, the neural fingerprints were generated with code from Duvenaud et al\cite{duvenaud_convolutional_2015}. The neural network used for prediction had one hidden layer of 100 units. Hyperopt \cite{Bergstra_2015} in conjunction with Scikit-learn \cite{scikit-learn} was used to optimize the learning rate, the initial scale, and the fingerprint length for each of the molecules. 

For some reaction types, certain reagents or secondary reactants are required for that reaction. Thus, it is possible that the algorithm may learn to simply associate these components in the reaction with the corresponding reaction type. As a baseline test to measure the impact of the secondary reactant and the reagent on the prediction, we also performed the prediction with a modified fingerprint. For the baseline metric, the fingerprint representing the reaction was a one-hot vector representation for the 20 most common secondary reactants and the 30 most common reagents.\added{ That is, if one of the 20 most common secondary reactants or one of the 30 most common reagents was found in the reaction, the corresponding bits in the baseline fingerprint were turned on; if one of the secondary reactants or reagents was not in these lists, then a bit designated for 'other' reactants or reagents was turned on.} This combined one-hot representation of the secondary reactants and the reagents formed our baseline fingerprint. 

Once a reaction type has been predicted by the algorithm, the SMARTS transformation associated with the reaction type is applied to the reactants. If the input reactants met the requirements of the SMARTS transformation, the products molecules generated by the transformation is the predicted structure of the products. If the reactants do not match the requirements of the SMARTS transformation, the algorithm instead guesses the structure of the reactants instead, i.e. it is assumed that no reaction occurs. 

\begin{acknowledgement}

The authors thank Rafael Gomez Bombarelli, Jacob Sanders, Steven Lopez, and Matthew Kayala for useful discussions. J.N.W. acknowledges support from the National Science Foundation Graduate Research Fellowship Program under Grant No. DGE-1144152. D.D. and A.A.-G. thank the Samsung Advanced Institute of Technology for their research support. The authors thank FAS Research computing for their computing support and computer time in the Harvard Odyssey computer cluster. All opinions, findings and conclusions expressed in this paper are the authors' and do not necessarily reflect the policies and views of NSF or SAIT.
\end{acknowledgement}

\begin{suppinfo}
The supporting information section contains SI Table 1: Comparison of similarity of exam question reactants and reagents to training set reactants and reagents. The code and full training datasets will be made available at \url{https://github.com/jnwei/neural_reaction_fingerprint.git}.
\end{suppinfo}

\bibliography{biblio}


\end{document}